\documentclass[8.5pt]{article}
\usepackage[super,sort&compress,comma]{natbib} 
\usepackage{mhchem}
\usepackage{times,mathptmx}
\usepackage{sectsty}
\usepackage{balance} 

\usepackage{eps fig} 
\usepackage{lastpage}
\usepackage[format=plain,justification=raggedright,singlelinecheck=false,font=small,labelfont=bf,labelsep=space]{caption} 
\usepackage{fancyhdr}
\newcommand{\etalp}{\textit{et~al.~}}

\begin{document}

\thispagestyle{plain}
\fancypagestyle{plain}{
\fancyhead[L]{\includegraphics[height=8pt]{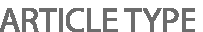}}
\fancyhead[C]{\hspace{-1cm}\includegraphics[height=20pt]{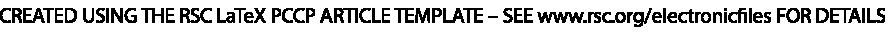}}
\fancyhead[R]{\includegraphics[height=10pt]{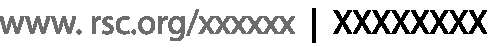}\vspace{-0.2cm}}
\renewcommand{\headrulewidth}{1pt}}
\renewcommand{\thefootnote}{\fnsymbol{footnote}}
\renewcommand\footnoterule{\vspace*{1pt}%
\hrule width 3.4in height 0.4pt \vspace*{5pt}} 
\setcounter{secnumdepth}{5}

\makeatletter 
\def\subsubsection{\@startsection{subsubsection}{3}{10pt}{-1.25ex plus -1ex minus -.1ex}{0ex plus 0ex}{\normalsize\bf}} 
\def\paragraph{\@startsection{paragraph}{4}{10pt}{-1.25ex plus -1ex minus -.1ex}{0ex plus 0ex}{\normalsize\textit}} 
\renewcommand\@biblabel[1]{#1}            
\renewcommand\@makefntext[1]%
{\noindent\makebox[0pt][r]{\@thefnmark\,}#1}
\makeatother 
\renewcommand{\figurename}{\small{Fig.}~}
\sectionfont{\large}
\subsectionfont{\normalsize} 

\fancyfoot{}
\fancyfoot[LO,RE]{\vspace{-7pt}\includegraphics[height=9pt]{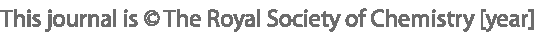}}
\fancyfoot[CO]{\vspace{-7.2pt}\hspace{12.2cm}\includegraphics{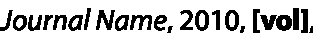}}
\fancyfoot[CE]{\vspace{-7.5pt}\hspace{-13.5cm}\includegraphics{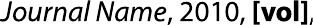}}
\fancyfoot[RO]{\footnotesize{\sffamily{1--\pageref{LastPage} ~\textbar  \hspace{2pt}\thepage}}}
\fancyfoot[LE]{\footnotesize{\sffamily{\thepage~\textbar\hspace{3.45cm} 1--\pageref{LastPage}}}}
\fancyhead{}
\renewcommand{\headrulewidth}{1pt} 
\renewcommand{\footrulewidth}{1pt}
\setlength{\arrayrulewidth}{1pt}
\setlength{\columnsep}{6.5mm}
\setlength\bibsep{1pt}

\twocolumn[
  \begin{@twocolumnfalse}
\noindent\LARGE{\textbf{Thermotropic interface and core relaxation dynamics of \mbox{liquid crystals} in silica glass nanochannels: A dielectric spectroscopy study}}
\vspace{1.6cm}

\noindent\large{\textbf{Silvia Ca{\l}us,\textit{$^{a}$} Lech Borowik,\textit{$^{a}$} Andriy V. Kityk,$^{\ast}$\textit{$^{a}$} Manfred Eich, \textit{$^{b}$} Mark Busch\textit{$^{c}$} and
Patrick Huber$^{\ast}$\textit{$^{c}$}}}\vspace{0.5cm}



\noindent \normalsize{We report dielectric relaxation spectroscopy experiments on two rod-like liquid crystals of the cyanobiphenyl family (5CB and 6CB) confined in tubular nanochannels with 7 nm radius and 340 micrometer length in a monolithic, mesoporous silica membrane. The measurements were performed on composites for two distinct regimes of fractional filling: monolayer coverage at the pore walls and complete filling of the pores. For the layer coverage a slow surface relaxation dominates the dielectric properties. For the entirely filled channels the dielectric spectra are governed by two thermally-activated relaxation processes with considerably different relaxation rates:  A slow relaxation in the interface layer next to the channel walls and a fast relaxation in the core region of the channel filling. The strengths and characteristic frequencies of both relaxation processes have been extracted and analysed as a function of temperature. Whereas the temperature dependence of the static capacitance reflects the effective (average) molecular ordering over the pore volume and is well described within a Landau-de Gennes theory, the extracted relaxation strengths of the slow and fast relaxation processes provide an access to distinct local molecular ordering mechanisms. The order parameter in the core region exhibits a bulk-like behaviour with a strong increase in the nematic ordering just below the paranematic-to-nematic transition temperature $T_{PN}$  and subsequent saturation during cooling. By contrast, the surface ordering evolves continuously with a kink near $T_{PN}$. A comparison of the thermotropic behaviour of the monolayer with the complete filling reveals that the molecular order in the core region of the pore filling affects the order of the peripheral molecular layers at the wall.
}
\vspace{0.5cm}
 \end{@twocolumnfalse}
  ]
\section{Introduction}

\footnotetext{\textit{$^{a}$~Faculty of Electrical Engineering, Czestochowa University of Technology, 42-200 Czestochowa, Poland, E-mail: andriy.kityk@univie.ac.at}}
\footnotetext{\textit{$^{b}$~Institute of Optical and Electronic Materials, Hamburg University of Technology (TUHH), D-21073 Hamburg-Harburg, Germany}}
\footnotetext{\textit{$^{c}$~Institute of Materials Physics and Technology, Hamburg University of Technology (TUHH), D-21073 Hamburg-Harburg, Germany, E-mail: patrick.huber@tuhh.de}}

Composites of liquid crystals (LCs) and monolithic, mesoporous solids, optically transparent porous silica in particular, are promising hybrid materials for organic electronics and the emerging field of nano photonics \cite{Martin1994, Schmidt-Mende2001, Bisoyi2011, Abdulhalim2012, Duran2012, Kumar2014}. They can be easily prepared by melt infiltration, profit from the mechanical stability of the porous solid template and from the large variation in electrical and optical properties offered by the plethora of different liquid crystalline/mesoporous host combinations nowadays available. Moreover, they allow one to explore the thermodynamics, structure and transport characteristics of liquid crystalline systems in restricted geometries, and thus phenomenologies which are of high interest both in nanoscience and nanotechnology \cite{Huber2015}.

Most prominently, the paranematic phase has been widely explored in theoretical \cite{Sheng1976, Poniewierski1987, Kutnjak2003, Kutnjak2004, Karjalainen2013, Karjalainen2015} and experimental\cite{Yokoyama1988, Bellini1992, Kralj1998, Kityk2008, Schoenhals2010, Grigoriadis2011, Calus2014} studies at planar surfaces and in porous media. It is characterised by a residual nematic order and thus the absence of a ''true'' isotropic liquid state at high temperatures. The evolution of the orientational order parameter from this pre-ordered state to the nematic phase can be described by a ''nematic ordering field'', $\sigma$ within a Landau-de-Gennes free energy approach \cite{Sheng1976, Poniewierski1987, Kutnjak2003, Kutnjak2004}. The strong first order \textit{I-N} transition is replaced by a weak first order or continuous paranematic-to-nematic (\textit{P-N}) transition at a temperature $T_{PN}$ and may also be accompanied by pre-transitional phenomena in the molecular orientational distribution \cite{Eich1984}. For tubular pore geometry, this effective field is strongly dependent both on the average pore radius $R$ ($\propto R^{-1}$) and on the LC-wall interaction.
 
Optical polarimetry \cite{Kityk2008,Kityk2010,Calus2012,Calus2014, Huber2015} provides arguably the most accurate insights in the order parameter behaviour in the vicinity of the paranematic-to-nematic transition \cite{Kutnjak2003,Kutnjak2004}. Note, however, that this technique and most other experimental methods probe an effective (averaged) molecular ordering. 

By contrast computer simulations on LCs in thin film and pore geometry can give spatially resolved information \cite{Gruhn1997, Gruhn1998, Barmes2004, Care2005, Binder2008, Ji2009, Ji2009b, Mazza2010, Pizzirusso2012, Roscioni2013,Karjalainen2013, Cetinkaya2013, Schulz2014, Karjalainen2013, Busselez2014, Karjalainen2015}. These studies indicate pronounced spatial heterogeneities, encompassing interface-induced molecular layering and radial gradients both in the orientational order and reorientational dynamics \cite{Li2009, Mazza2010}. 


\begin{figure}[tbp] \center
\epsfig{file=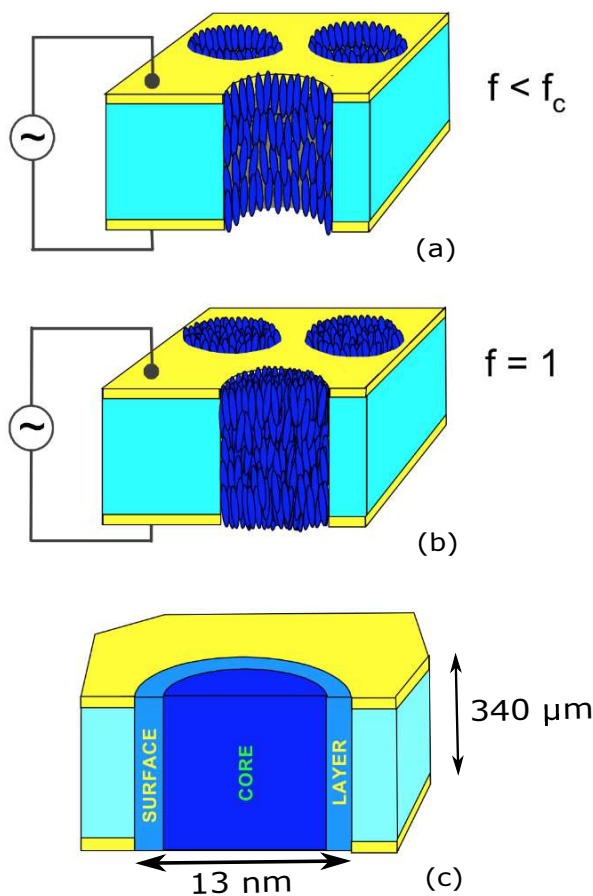, angle=0, width=0.9\columnwidth}\caption{Sketch of the capacitor geometry for the dielectric relaxation experiments: Gold electrodes were evaporated onto the mesoporous silica membrane, so that the sample forms a simple parallel circuit both in the regime of monomolecular layer coverage (a) and for the completely filled channels (b). (c) Illustration of two regions with two distinct molecular mobility for completely filled nanochannels: the interface layer next to the channel wall with a slow dipolar relaxation and the core region characterised by fast relaxation dynamics.} \label{fig1}
\end{figure}

Dielectric relaxation spectroscopy has proven as particularly suitable to achieve such local information on confined LCs \cite{Cramer1997, Hourri2001, Frunza2001, Leys2005, Sinha2005, Leys2008, Frunza2008, Bras2008, Jasiurkowska2012, Calus2015}. A number of studies on LCs in porous media evidently document that the rate of dipolar relaxations (and thus orientational and translational mobility) usually differ significantly between the molecules in the pore wall proximity and the ones in the channel centre \cite{Cramer1997, Hourri2001, Frunza2001, Leys2005, Sinha2005, Leys2008, Frunza2008, Bras2008, Jasiurkowska2012}. 

In dielectric studies Aliev \etalp \cite{Sinha1997, Sinha1998,Aliev2005} found a slow surface mobility in comparison to the dynamics in the pore centre for liquid crystals confined in tortuous and tubular mesopores. Moreover, a significant broadening of the dielectric spectra was observed and traced by the authors to inhomogeneous couplings of the molecules to the pore walls and coupling variations among the molecules themselves \cite{Aliev2010}. 

It turned out that the rate of dipolar relaxation differs by about two orders of magnitude between the molecules located at the host-guest interface and the ones located in the core region of the pore filling, see sketch in Fig.~1c. Therefore, both relaxation processes can be discriminated and the corresponding dipolar relaxation strengths provide then access to local molecular ordering. Employing this distinct dynamical behaviour, we could recently show for 7CB, a prominent member of the rod-like cyanobiphenyl nematogens that the ordering in the core region is reminiscent of the bulk behaviour \cite{Calus2015}. It is characterised by an abrupt increase in the nematic ordering in the transition region. By contrast, the surface ordering exhibits a continuous thermotropic evolution of nematic order with a gradual change in slope and an asymptotic vanishing in the paranematic phase.

In this article we extend our previous dielectric spectroscopy studies on two other members of the cyanobiphenyl family, i.e. $n$CB ($n$=5,6) embedded in silica membranes with parallel aligned nanochannels. Overall, we find a very similar behaviour of the effective average as well as local order parameters as in the case of 7CB. As we will outline below, it is again possible to describe the experimental observations in a phenomenological manner by applying a Landau-De Gennes model. The analogous findings and semi-quantitative descriptions document that the dielectric spectroscopy technique along with the phenomenological approach and key physical principles outlined in Ref. \cite{Calus2015} seem to be applicable to confined nematics in general.

\section{Experimental}

The nematic LCs 5CB and 6CB have been purchased from Merck, Germany. The porous silica ($p$SiO$_2$) membranes were obtained by electrochemical anodic etching of highly $p$-doped  $\langle$100$\rangle$ silicon wafers which have been subjected to thermal oxidation for 12 h at $T$=800 $^o$C under standard atmosphere. The resulting array of channels are aligned along the [100] crystallographic direction, i.e. perpendicular to the membrane surface. The average channel radius is $R=6.6 \pm$0.5 nm (porosity $P=55 \pm$2\%) as determined by recording volumetric N$_2$-sorption isotherms at $T$=77~K. For the dielectric measurements, gold electrodes have been deposited onto the porous membrane. All measurements were performed on samples cut from a monolithic porous membrane of $d=$340 $\mu$m thickness. Two samples with electrode area of 97 mm$^2$ (geometric capacitance $C_0=\varepsilon_0 S/d=$ 2.52 pF) and 104 mm$^2$ ($C_0=$ 2.71 pF) have been filled by nematic LCs 5CB and 6CB, respectively.

The entirely filled samples (fractional filling $f=$1.0) have been obtained by capillary imbibition of the LC melt \cite{Gruener2011}. To prepare the partially filled samples with layer coverage we immersed it into a binary LC/cyclohexane solution (5 vol. \%) for about 20-30 minutes. After evaporation of the high-vapour-pressure solvent (cyclohexane), the low-vapour-pressure LC remained in the porous matrix. The final filling fraction $f$ and the complete evaporation of the solvent have been verified by comparing the sample weight before and during the adsorption procedure \cite{Huber2013}. The obtained fraction fillings $f=0.16\pm0.01$ (for 5CB in $p$SiO$_2$) and $f=0.18\pm0.01$ (for 6CB in $p$SiO$_2$) correspond approximately to one monomolecular layer.

\begin{figure*}[htbp] \center
\epsfig{file=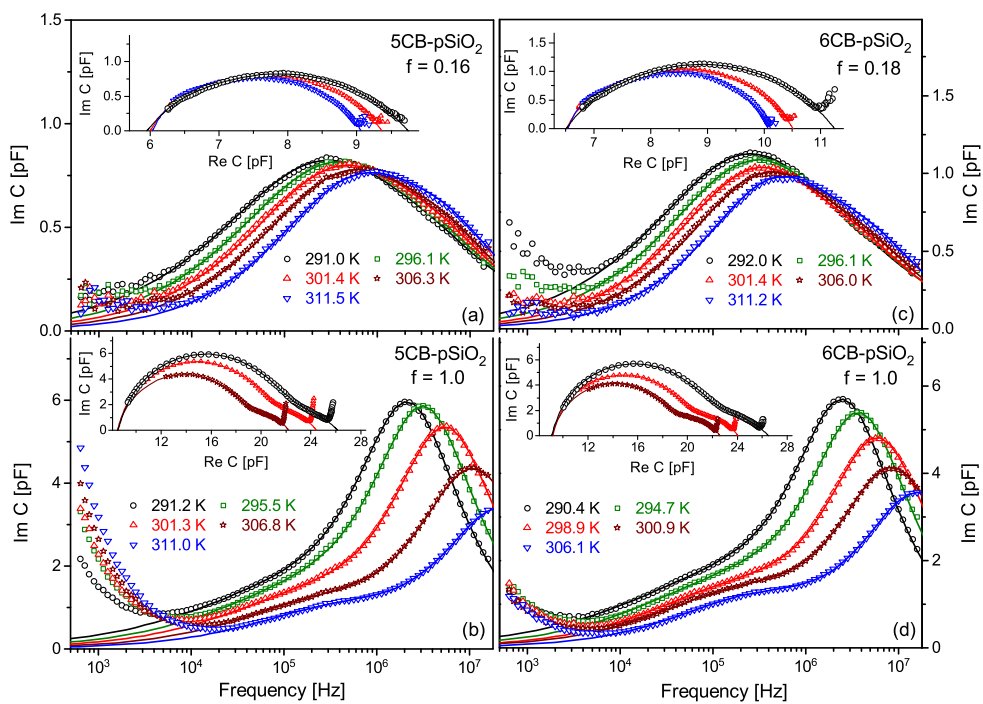, angle=0,width=1.5\columnwidth} \caption{\footnotesize Frequency dispersion of the imaginary capacitance $C''(\nu)$ in 5CB-$p$SiO$_2$ and 6CB-$p$SiO$_2$ nanocomposites for five selected temperatures. Panels (a) and (c) correspond to partially filled matrices 5CB-$p$SiO$_2$ ($f=0.16$) and 6CB-$p$SiO$_2$ ($f=0.18$), respectively, in the regime of monomolecular layer coverage. Panels (b) and (d) correspond to nanocomposites with entirely filled pores ($f=1.0$). Insets in panels (a)-(d) displays Cole-Cole plots for three selected temperatures. Symbols in the panels and inserts are the measured data points. Solid lines are the best fits based on the Cole-Cole relaxation model (see Eq.(1)). The geometric capacitance, $C_0$, equals 2.52 pF (5CB-$p$SiO$_2$) and 2.71 pF (6CB-$p$SiO$_2$).}
 \label{fig2}
\end{figure*}

Dielectric spectra have been recorded in the frequency range from 0.5 kHz to 15 MHz using the impedance/gain-phase analyser Solatron-1260A. The measurements have been performed at selected temperatures between 292~K and 324~K  and controlled with an accuracy of 0.01~K by a LakeShore 340 temperature controller.

Note that both, the entirely filled samples and the partially filled samples with monolayer coverage form simple parallel electrical circuits, as schematically sketched in Figs.~1a and 1b, respectively. The advantage of the simple geometry is evident: Since the channels are all arranged parallel to the external electric field, the effective complex permittivity of the composite, $\varepsilon^*$, is given by the permittivities of the components $\varepsilon^*_{SiO_2}$ = 3.8 (silica substrate), $\varepsilon^*_{nCB}$ (nematic LC) and $\varepsilon_{v}$ = 1 (vacuum permittivity) weighted with the corresponding volume fractions: $\varepsilon^* = (1-P)\varepsilon^*_{SiO_2} + P [f\varepsilon^*_{nCB} + (1-f)\varepsilon_{v}]$. In the frequency range 0.0005-15 MHz $\varepsilon^*_{SiO_2}\approx \varepsilon'_{SiO_2}$ and and it does not exhibit any frequency dispersion. Thus, the relaxation behaviour observed in the effective permittivity, $\varepsilon^*(\nu)$ correspond to that of the confined LC. For more complicated pore geometries, e.g. for partial fillings in the capillary condensation regime depolarisation effects may lead to phase shifts of the internal electric field compared to the external field, which considerably complicates the analysis - see also Ref.~\cite{Calus2015}. Of course, the channel walls of the present matrix are rough, and the radius is likely to vary along the long channel axis which may result in deviations from the ideal parallel geometry. Nevertheless, the parallel-circuit assumption is expected to be a good approximation.

\section{Results and discussion}
\subsection{Molecular mobility probed by dielectric spectroscopy}

In Fig.~2 we display the frequency dispersion of the imaginary capacitance $C''(\nu)=\varepsilon''(\nu)C_0$ (symbols) at five selected temperatures, $T$, for entire and partially filled nanoporous silica membranes. 

In order to analyse the observed relaxation behaviour in more detail, we resort to a representation of the dielectric data in so-called Cole-Cole diagrams, i.e. we plot the imaginary part $C''(\nu)$ on the vertical axis and the real part $C''(\nu)$ on the horizontal axis with frequency $\nu$ as the independent parameter, see insets in Fig.~2. In such a diagram, a material that has a single relaxation frequency, as typical of the classical Debye relaxator, will appear as a semicircle with its center lying on the horizontal at $C=0$ and the peak of the loss factor occurring at 1/$\tau$, where $\tau$ is a measure of the mobility of the molecules (dipoles). It characterises the time required for a displaced system aligned in an electric field to return to $1/e$ of its random equilibrium value (or the time required for dipoles to become oriented in an electric field). A material with multiple relaxation frequencies will be a semicircle (symmetric distribution) or an arc (nonsymmetrical distribution) with its center lying below the horizontal at $C=0$. In that sense, the Cole-Cole representation allows one to geometrically illustrate and analyse the relaxation behaviour of a given system in a quite simple manner. \cite{Cole1941, Kremer2002}
An analysis of the experimental data shows that the dielectric dispersion of the complex capacitance $C^*(\omega)$ can be well described by two Cole-Cole processes:
\begin{eqnarray}
C^*(\omega)&=&\varepsilon^*(\omega) C_0=\nonumber \\
&=&C_{\infty}+\frac{\Delta C_1}{1+(\textrm{i}\omega \tau_1)^{1-\alpha_1}}+\frac{\Delta C_2}{1+(\textrm{i}\omega \tau_2)^{1-\alpha_2}}, \quad
\label{eq1}
\end{eqnarray}
where $\omega=2\pi\nu$ is the cyclic frequency,  $C_{\infty}=\varepsilon_{\infty}C_0$ is the high frequency limit capacitance expressed via the high frequency permittivity $\varepsilon_{\infty}$, $\Delta C_1=\Delta \varepsilon_1 C_0$ and $\Delta C_2=\Delta \varepsilon_2 C_0$ are the capacitance relaxation strengths expressed via the dielectric relaxation strengths $\Delta\varepsilon_1$ and $\Delta \varepsilon_2$ of the slow process I and the fast process II, respectively, $\tau_1$ and $\tau_2$ are the mean relaxation times of the corresponding processes. The Cole-Cole exponents $\alpha_1$ and $\alpha_2$ ($0\le\alpha_1,\alpha_2<1$) characterise the distribution of the relaxation times; its zero limit corresponds to the Debye process with a single relaxation time. Solid lines in Fig.~2 are the best fits as obtained by a simultaneous analysis of the measured real and imaginary capacitance based on Eq.~1. The deviations of the experimental data points from the fitting curves at low frequencies originate in ionic dc-conductivity. For this reason the low-frequency region was always excluded from the fitting analysis.This fitting procedure has been applied to the measured dispersion curves in the entire temperature range. In the case of entirely filled samples, 5CB-$p$SiO$_2$ and 6CB-$p$SiO$_2$ ($f=1.0$), uncertainties regarding the extracted fit parameters can occur at higher temperatures, particularly above the paranematic-to-nematic transition point, $T_{PN}$. Then the maximum of the imaginary part, $C''(\nu)$, shifts out of the upper limit of the frequency window ($\nu_m>$15 MHz) employed in the dielectric measurements. Thus only a part of the left wing of the relaxation band is observed. 

Fortunately, this problem can be resolved by following the route described in detail in Ref.\cite{Calus2015}, where one takes into account that the capacitance relaxation strength  $\Delta C_2$ can be determined alternatively by employing Eq. (1) in its static limit ($\nu \rightarrow 0$):
\begin{equation}
\Delta C_2 =C_{\mathrm{st}}-C_{\infty}-\Delta C_1.
\label{eq2}
\end{equation}

Here the static dielectric constant, $C_{\mathrm{st}}$, is determined directly from the Cole-Cole plot by its extrapolation to the low frequency region. Obviously, $C_{\mathrm{st}}$ is practically independent from other extracted fit parameters. $C_{\infty}$, on the other hand, it is strongly dominated by the electronic polarisability which is weakly temperature dependent. Estimations show that its contribution to the temperature changes of the static capacitance, similarly as in the case of 7CB-$p$SiO$_2$ \cite{Calus2015}, do not exceed 5\% in the entire $T$-range, i.e. it remains within the error of the fitting analysis. In the fitting procedure the best parameter sets have been determined by a minimalisation of the difference between the $\Delta C_2$ values extracted by the two alternative ways: (i) direct fits of measured dispersion $C'(\nu)$ and $C''(\nu)$ and (ii) extracting $C_{\mathrm{st}}$-value from the Cole-Cole plots and subsequent employing of Eq.(2). The combination of these procedures ensures unambiguity of the extracted fit parameters. The corresponding temperature dependences are displayed in Figs.~3-6.

The interpretation of the dispersion curves is similar to that reported in Ref.\cite{Calus2015} as well as in a series of previous dielectric studies \cite{Hourri2001,Leys2005,Sinha2005,Leys2008,Bras2008,Kityk2014,Wallacher2004b}. For the entirely filled samples ($f=1$) the slow relaxation corresponds to the rotational dynamics of the molecules in direct contact with the pore walls (so-called surface or more strictly spoken interface relaxation), see sketch in Fig.~1c. The fast relaxation process originates in the molecular rotational dynamics in the core region (the so-called core relaxation). It is obvious that for partially filled samples, i.e. in the regime of monomolecular layer coverage, the dielectric relaxation is strongly dominated by the surface relaxation. 

Nevertheless, a weak contribution of the fast relaxation is also present. Its presence is indicated by the slightly asymmetric shape of Cole-Cole plots observed at higher temperatures. The corresponding relaxations strength, $\Delta C_2$ (see Fig.~6, $f=0.16$ (5CB) and $f=0.18$ (6CB)), decrease with decreasing temperature and practically vanish below 290~K. We believe that due to increasing intermolecular spacings at higher temperatures some molecules are pushed into the next molecular layer, resulting in a considerably faster dipolar relaxation. The ''fast'' relaxation process in the monomolecular layer regime, on the other hand, differs qualitatively from the one observed in the core region of entirely filled matrices. In the first case the molecules are located at the tbp interface. In the second one they are located in the core region of the pore filling and, accordingly, are influenced by the collective molecular ordering resulting from the paranematic-to-nematic transition. Therefore, the smooth temperature variations of all Cole-Cole parameters observed in the monomolecular layer regime contrast with the ones observed for the entirely filled matrices. Roughly speaking, the monomolecular layer regime does not show a phase transition. This is evident if one compares e.g. the temperature dependences of the static capacitance, $C_{\mathrm{st}}(T)$, for both regimes of pore filling, see Fig.~5.

\begin{figure}[tbp]
\begin{center} \center
\epsfig{file=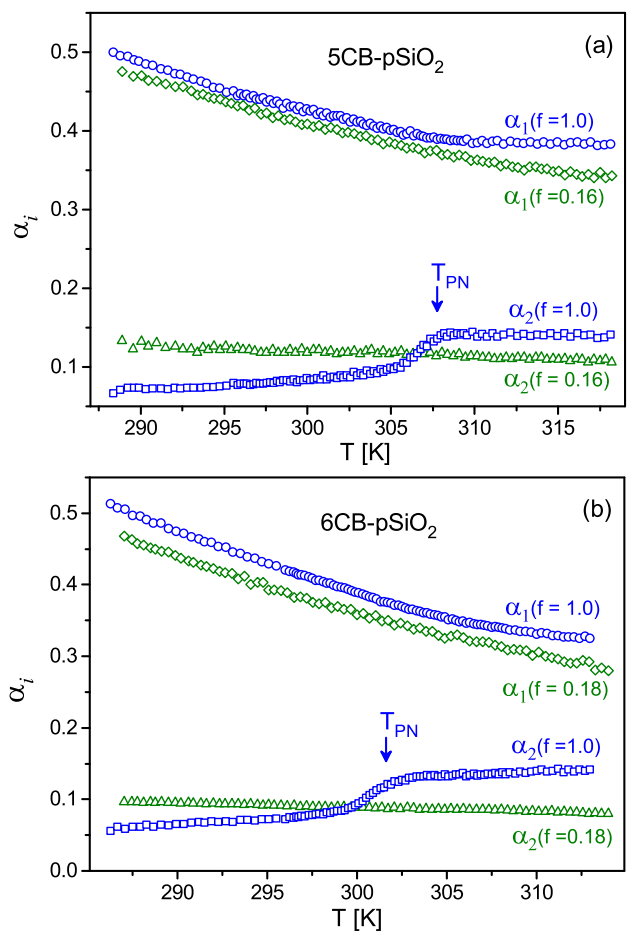, angle=0, width=0.99\columnwidth}
\caption{\footnotesize Cole-Cole parameters of the slow relaxation process I, $\alpha_1$, and fast relaxation process II, $\alpha_2$  vs $T$ as extracted within the fitting procedure for 5CB-$p$SiO$_2$ (a)  and 6CB-$p$SiO$_2$ (b) nanocomposites in the regimes with entirely ($f=1.0$) and partial ($f=0.16$ (5CB-$p$SiO$_2$); $f=0.18$ (6CB-$p$SiO$_2$)) filled channels.} \label{fig3}
\end{center}
\end{figure}

\begin{figure}[tbp]
\begin{center} \center
  \epsfig{file=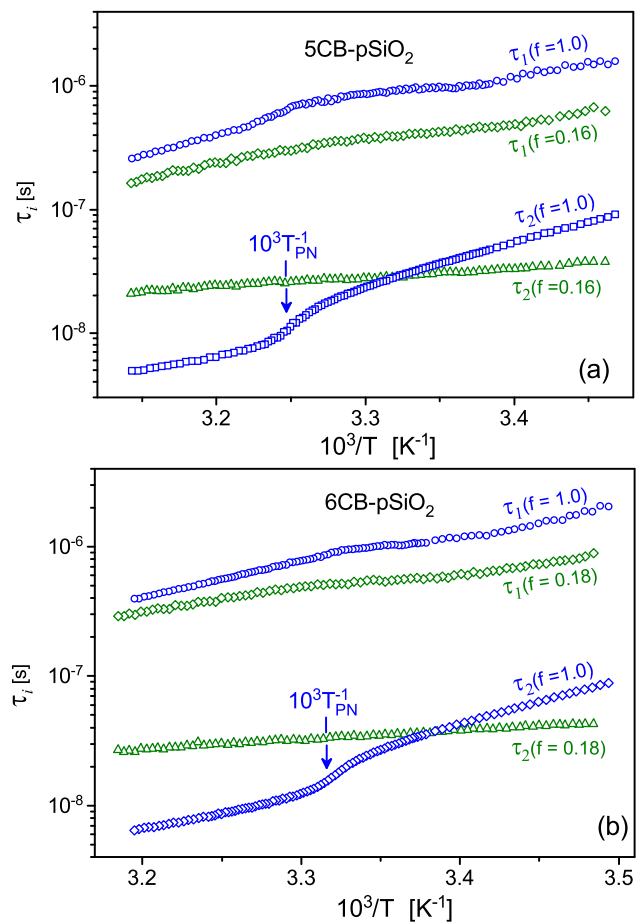, angle=0, width=0.99\columnwidth}
  \caption{\footnotesize Relaxation times of the slow relaxation process I, $\tau_1$, and  fast relaxation process II, $\tau_2$  vs $T^{-1}$ as extracted within the fitting procedure for 5CB-$p$SiO$_2$ (a)  and 6CB-$p$SiO$_2$ (b) nanocomposites in the regimes with entirely ($f=1.0$) and partial ($f=0.16$ (5CB-$p$SiO$_2$); $f=0.18$ (6CB-$p$SiO$_2$)) filled channels.} \label{fig4}
\end{center}
\end{figure}

For the nanocomposites with entirely filled matrices ($f=$1.0) the Cole-Cole parameters, $\alpha_1$ and $\alpha_2$ (see Fig.~3) exhibit opposite tendencies in their temperature evolution. The surface relaxation (process I) is characterised by a rather broad distribution of the relaxation times at room temperature  ($\alpha_1 \sim$ 0.48-0.50), but becomes somewhat narrower at high temperatures and approaches a magnitude of 0.35-0.40 in the paranematic phase. The dipolar relaxation in the core region (process II), on the other hand, is characterised by a narrow distribution of the relaxation rates at room temperature ($\alpha_2 \sim$ 0.05-0.06) and rises to about 0.12-0.14 in the paranematic state, with a characteristic, step-like change in the vicinity of the paranematic-to-nematic transition. Similar changes in the phase transition region are observed in the temperature evolution of other relaxation parameters, $\tau_2(T)$ (Fig.~4) and $\Delta C_2(T)$ (Fig.~6). This means that an orientational ordering in the core region affects the relaxation parameters. For comparison, the surface relaxation parameters, $\alpha_1$, $\tau_1$ and $\Delta C_1$ exhibit here only smooth temperature variations. At this stage, one can already conclude that the molecular ordering in the core and surface regions of the pore filling are expected to be considerably different. This issue will be discussed in details below.

For the pores with monomolecular layer coverage, the parameter $\alpha_1$ exhibits quite similar temperature variation as the one found for the entirely filled samples. The corresponding $\alpha_1(T)$-curves are slightly shifted down, see Fig.~3.

\begin{figure}[tbp]
\begin{center} \center
  \epsfig{file=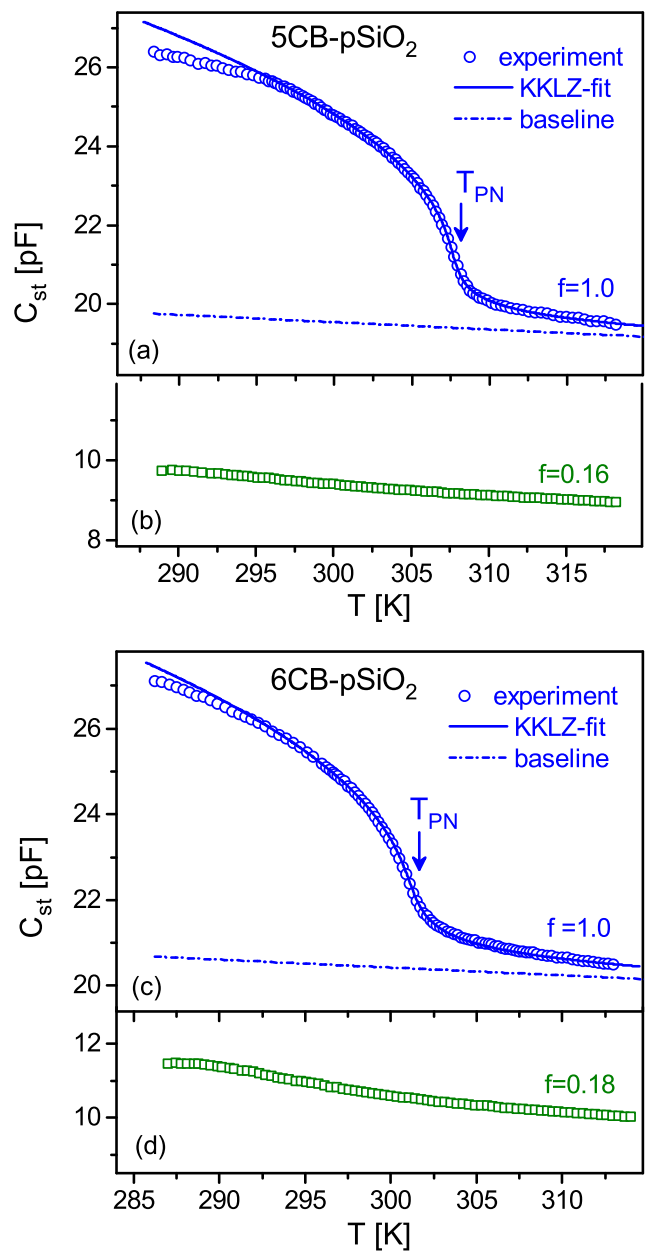, angle=0, width=0.99\columnwidth}
  \caption{\footnotesize Static capacitance, $C_{\mathrm{st}}$ as extracted within the fitting procedure (symbols) for 5CB-$p$SiO$_2$ ($f=1.0$(a), $f=0.16$(b)) and 6CB-$p$SiO$_2$ ($f=1.0$(c), $f=0.18$(d)) nanocomposites. Solid blue lines are the best fits of $C_{\mathrm{st}}(T)$-dependence ($f=$1.0) based on the KKLZ approach. Dash-dot blue lines indicate isotropic baselines, $C_t^{\mathrm{(iso)}}(T)$, obtained within the same fitting procedure.} \label{fig5}
\end{center}
\end{figure}

The relaxation times of the surface and core processes are displayed in Fig.~4. The fast relaxation in the core region exhibits Arrhenius-like behaviour ($\tau_2=\tau_{o2}\exp(E_{a2}/k_bT)$) with somewhat different activation energies in the nematic ($T << T_{PN}$) and paranematic ($T >> T_{PN}$) phases. An exception is the vicinity of the paranematic-to-nematic transition, where smeared, step-like variations are observed. They may be attributable to a specific behaviour of the attempt relaxation time, $\tau_{o2}$, presumably because of changes in the activation entropy caused by the orientational molecular ordering at the phase transformation. The surface relaxation, $\tau_1$, on the other hand, exhibits only a smooth temperature variation accompanied by a gradual change in slope, but with no special features in the vicinity of $T_{PN}$. This process is much slower because of the large viscosity in the surface layer compared to the bulk viscosity. This hinders the rotational dynamics of the molecules \cite{Sinha1998}. Both relaxation times rise with increasing orientational ordering which is typical of rotational dynamics around the short molecular axis \cite{Haws1989,Diez2006}.

\subsection{Thermotropic nematic order: Landau-de Gennes Analysis and Surface/Core Partitioning}

In the Landau-de~Gennes theory the nematic (orientational) order parameter is defined as $Q=\frac{1}{2}\langle3\cos^2\theta-1\rangle$, where $\theta$ is the angle between the axis of the molecule and the director, $\vec{n}$, and brackets mean an averaging over the ensemble of molecules. The size of an ensemble domain is an important issue in such a description, particularly when the order parameter is spatially inhomogeneous. Smaller sizes are appropriate to describe local properties of the molecular ordering. Larger ones are more suitable in a characterisation of an effective (averaged) value of the order parameter. Its appropriate choice is thus a matter of the specific system under consideration and/or the approach used for its description. Specifically, the surface layer and core region of the systems studied here may be considered as two regions with homogeneous order behaviour, characterised by two distinct, local order parameters, $Q_s$ and $Q_c$, respectively.
The effective (averaged) order parameter over the pore filling, $\bar{Q}$, can then be expressed as a superposition of the elementary contributions:  $\bar{Q}=\bar{Q_s}+ \bar{Q_c}$, where $\bar{Q_s}=w_sQ_s$ and $\bar{Q_c}=w_cQ_c$, $w_s$ and $w_c$ ($w_s+w_c=1$) are the weight factors (volume fractions) of surface and core components, respectively. Although quite simplistic, this approach has an evident practical benefit: All three quantities, i.e. $\bar{Q}$, $\bar{Q_s}$ and $\bar{Q_c}$  can be extracted independently from the experiment. Following the approach developed in Ref.\cite{Calus2015}, which is based on the Maier and Meier equation \cite{Maier1961} the excess changes of the static capacitance are proportional to $\bar{Q}$:
\begin{equation}
C_{\mathrm{st}}-C_t^{\mathrm{(iso)}}\propto \bar{Q},\label{eq3}
\end{equation}
where $C_t^{\mathrm{(iso)}}(T)$ is the \emph{isotropic baseline} of the static capacitance. In the case of the bulk nematic  LC $\Delta C_t^{\mathrm{(iso)}}(T)$ represents the bare temperature dependence of the capacitance relaxation strength in the isotropic phase, whereas its extrapolation to the nematic phase gives an isotropic baseline relative to which an excess contribution due to an orientational ordering ($p$-term) is counted. Similarly, an excess changes of the capacitance relaxation strengths $\Delta C_1$ and $\Delta C_2$ are proportional to $\bar{Q_s}$ and $\bar{Q_c}$, respectively:
\begin{eqnarray}
\Delta C_1(T) -  \Delta C_1^{\mathrm{(iso)}}(T)\propto \bar{Q_s}(T); \\ \nonumber
\Delta C_2(T) -  \Delta C_2^{\mathrm{(iso)}}(T)\propto \bar{Q_c}(T). \label{eq4}
\end{eqnarray}
where  $\Delta C_1^{\mathrm{(iso)}}(T)$ and $\Delta C_2^{\mathrm{(iso)}}(T)$ are the corresponding isotropic baselines. The factor of proportionality depends on the molecular dipole moment and its orientation with respect to the long principal axis of the molecule, the molecular number density, the internal field factors and the temperature, but it is identical in Eqs.~3 and 4.

\begin{figure}[tbp]
\begin{center}
  \epsfig{file=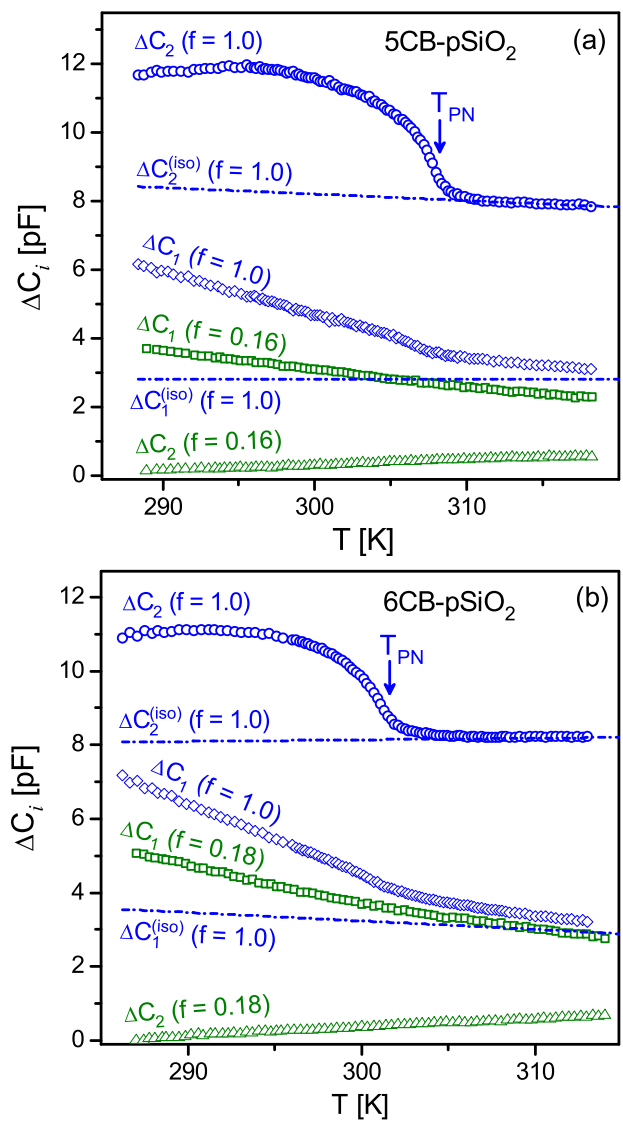, angle=0, width=0.99\columnwidth}
  \caption{\footnotesize Capacitance relaxation strengths of the slow relaxation process I, $\Delta C_1$ and fast relaxation process II, $\Delta C_2$  vs $T$ as extracted within the fitting procedure for 5CB-$p$SiO$_2$ (panel (a), $f=1.0$, $f=0.16$)  and 6CB-$p$SiO$_2$ (panel (b), $f=1.0$ $f=0.18$) nanocomposites. Dash-dot blue lines indicate the baselines of the isotropic state, see labels.} \label{fig6}
\end{center}
\end{figure}

In Fig.~5 we display the temperature dependences of the static capacitance for the composites 5CB-$p$SiO$_2$ and 6CB-$p$SiO$_2$. In the monomolecular layer regime [5CB-$p$SiO$_2$ ($f=0.16$), section (b); 6CB-$p$SiO$_2$ ($f=0.18$) section (d)]  only a gradual increase of $C_{\mathrm{st}}$ is observed upon cooling indicating a weak orientational ordering, but no hints of a phase transformation. For the entirely filed sample [5CB-$p$SiO$_2$ ($f=1.0$), section (a); 6CB-$p$SiO$_2$ ($f=1.0$) section (c)], on the other hand,  pronounced changes with a characteristic kink at $T_{PN}$ on $C_{\mathrm{st}}(T)$ dependences are observed for both nanocomposites. A typical feature of these dependences is their \emph{continuous} character with a precursor behaviour quite similar to that observed in recent optical birefringence studies \cite{Kityk2008,Kityk2010,Calus2012,Calus2014}.

\begin{figure*}[tbp]
\begin{center} \center
  \epsfig{file=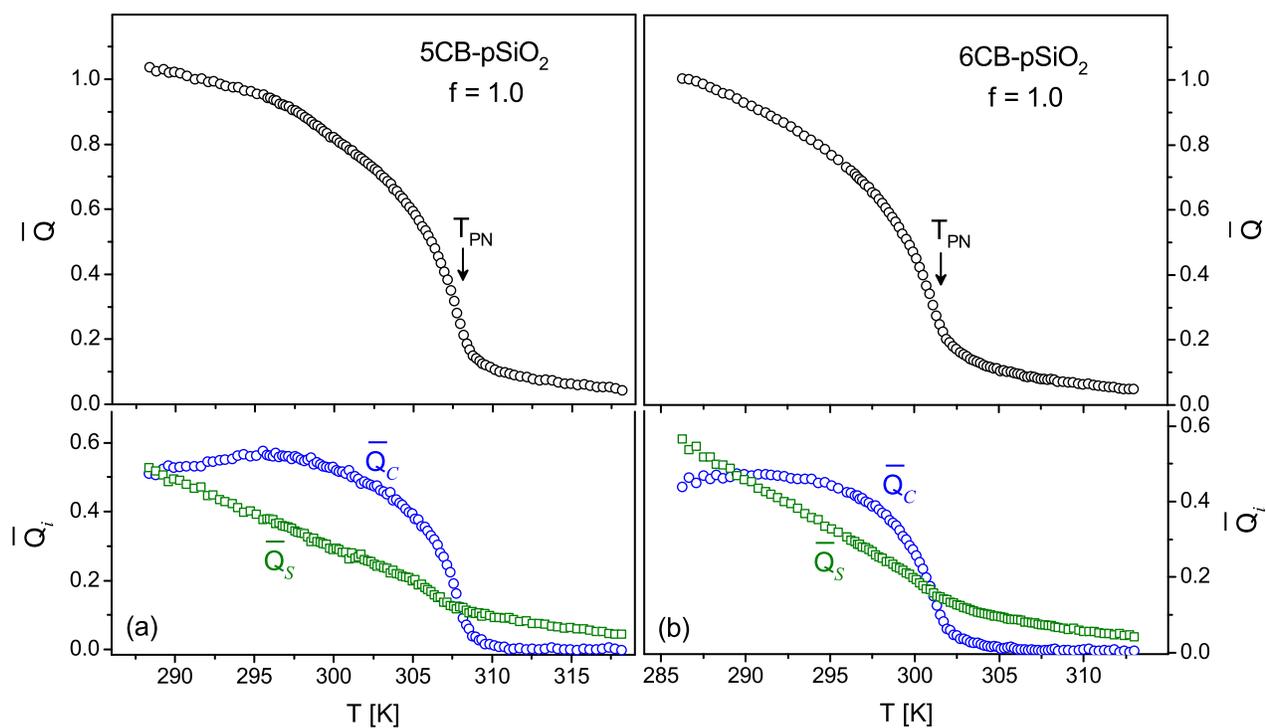, angle=0, width=1.99\columnwidth}
  \caption{\footnotesize  The effective (averaged) orientational order parameter, $\bar{Q}$ vs $T$ and its deconvolution in the elementary contributions (averaged local order parameters) describing molecular orderings in the surface ($\bar{Q_s}(T)$) and core ($\bar{Q_c}(T)$) regions for 5CB-$p$SiO$_2$ ($f=1.0$, section (a)) and 6CB-$p$SiO$_2$ ($f=1.0$, section (b)) nanocomposites. For convenience all order parameters are normalised to the value $\bar{Q}$ at 290~K. The relation $\bar{Q}=\bar{Q_s}+\bar{Q_c}$ holds for each $T$.}\label{fig7}
\end{center}
\end{figure*}

Whereas the molecular ordering in the bulk, i.e. the isotropic-to-nematic phase transition, can be described by a Landau-de Gennes theory, Kutnjak, Kralj, Lahajnar, and Zumer \cite{Kutnjak2004,Kutnjak2003} have extended the corresponding phenomenological approach (hereafter denoted as KKLZ model) towards nematic phases in cylindrical confinement. In principal, this model is applicable to entirely filled channels, only and its description depends on the anchoring conditions at the channel walls. The untreated silica channels enforce planar anchoring \cite{DrevensekOlenik2003} with a preferred orientation of LC molecules along the long channel axes. A key point of the KKLZ theory is a bilinear coupling between the order parameter and the nematic ordering field, $\sigma$, which results in a residual nematic ordering (paranematic state) for $T>T_{PN}$ instead of an isotropic one. Omitting a description of the KKLZ model and details of the fitting procedure, which can be find in Refs.\cite{Kityk2008,Kityk2010,Calus2012,Calus2014}, we present here only the main results of this analysis. Figs.~5a and~5c display the best fits of measured $C_{\mathrm{st}}(T)$-dependences along with the isotropic baselines, $C_t^{\mathrm{(iso)}}(T)$, obtained within the same fitting procedure. The difference between the bulk and effective transition temperatures, $\Delta T^*$, which calibrates the temperature scale of the KKLZ-model, was taken to be equal 3.4~K (5CB) and 3.2~K (6CB) based on the results of recent optical polarimetric studies \cite{Calus2012}. The best fits yield $\sigma$-values of 0.91 (5CB-$p$SiO$_2$) and 1.10 (6CB-$p$SiO$_2$), which corresponds to a supercritical regime in both cases. The magnitudes of the critical radius, $R_c=2R\sigma$ equal to 12.0~nm (5CB-$p$SiO$_2$) and 14.5~nm (6CB-$p$SiO$_2$), respectively, which within an experimental error ($\pm$1.0 nm) is in agreement with that obtained in the optical birefringence studies: 12.1~nm and 14.0~nm \cite{Calus2012}, respectively. Obviously, the KKLZ approach gives an adequate description of the paranematic-to-nematic transition. Deviations of the experimental data points from the fit curves observed at lower temperatures originate in order parameter saturation, which is not appropriately described by the free energy expansion of the KKLZ-model, since it is limited to a fourth-order expansion of the free energy.

In accordance with Eq.(3) the difference, $C_{\mathrm{st}}(T)- C_t^{\mathrm{(iso)}}(T)$ is proportional to $\bar{Q}(T)$. Normalised to a $\bar{Q}$-value at $T=290$ K, the $\bar{Q}(T)$-dependences of 5CB-$p$SiO$_2$ and 6CB-$p$SiO$_2$ nanocomposites are displayed in the upper panels of Figs.~7a and 7b, respectively. Using the $\Delta C_1(T)$ and $\Delta C_2(T)$ dependences presented in Fig.~6 for entirely filled matrices ($f=1.0$) and Eqs.(4)
the effective (averaged) order parameter $\bar{Q}$ can be deconvoluted into elementary contributions characterising molecular orderings in the surface ($\bar{Q_s}$) and core ($\bar{Q_c}$) regions. A principal challenge are the determination of the isotropic baselines, $\Delta C_i^{\mathrm{(iso)}}(T)$ ($i=1,2$). We trace first $\Delta C_2^{\mathrm{(iso)}}(T)$. Above $T_{PN}$ $\Delta C_2(T)$ rather fast saturates to a nearly temperature independent value. A linear extrapolation of this dependence below $T_{PN}$, as it is displayed in Fig.~6 represents the isotropic baseline in the confined nematic phase. This approach is not applicable, however, to describe the $\Delta C_1^{\mathrm{(iso)}}(T)$-baseline. $\Delta C_1(T)$ exhibits evident nonlinear asymptotic behaviour extending far above $T_{PN}$. Fortunately, having two other isotropic baselines it can be calculated as: $\Delta C_1^{\mathrm{(iso)}}(T)= C_t^{\mathrm{(iso)}}(T)-C_{\infty}-\Delta C_2^{\mathrm{(iso)}}(T)$, see labeled dash-dotted lines in Fig.~6. The difference $\Delta C_i(T)- \Delta C_i^{\mathrm{(iso)}}(T)$ ($i=1,2$) is proportional to elementary contributions, $\bar{Q_s}$ and $\bar{Q_c}$. Normalised again to $\bar{Q}$-value at ($T=$290 K), $\bar{Q_s}(T)$ and $\bar{Q_c}(T)$ dependences are displayed in the lower panels of Fig.~7a and 7b as for 5CB-$p$SiO$_2$ ($f=1.0$) and 6CB-$p$SiO$_2$ ($f=1.0$), respectively. Variations $\bar{Q_s}(T)$ and $\bar{Q_c}(T)$ in the region of $T_{PN}$ are considerably different. 

The molecular ordering in the core region is reminiscent of the evolution of the bulk order parameter. It exhibits a strong increase in the nematic ordering just below $T_{PN}$ and its subsequent saturation at $T<<T_{PN}$. The surface ordering, on the other hand, exhibits a continuous change with a smeared kink near $T_{PN}$ and an asymptotic decrease above this temperature. 
Moreover, the molecular ordering in the core region has an influence on the ordering in the periphery, particularly in the molecular layer located next to pore walls. This can be inferred by comparing the $\Delta C_1(T)$-dependences measured for the complete and partial pore fillings, see Fig.~6. Particularly,  a change in slope below $T_{PN}$ is quite obvious only for the samples with entirely filled pores. Presumably, the molecular ordering in the core region is transferred to the periphery via intermolecular interactions.

\section{Conclusion}

In conclusion, we reported a dielectric study on the calamatic nematic crystal 5CB and 6CB confined in cylindrical nanochannels of monolithic silica membranes. The measurements have been performed on composites with two distinct regimes of pore fillings: monomolecular layer coverages and entirely filled pores. Whereas for the composites with monomolecular layer coverages a slow surface relaxation dominates the dielectric properties, the dielectric spectra of nanocomposites with entirely filled pores can be well described by two Cole-Cole processes with well separated relaxation times. The fast relaxation is associated with a rotational dynamics of molecules in the core region of the pore filling. The slow relaxation originates from a surface LC layer next to the pore walls. 

In the regime of monomolecular layer coverage the temperature evolution of the static capacitance exhibits a smooth temperature behaviour with no hints of a phase transformation in the entire temperature region. This is in contrast to the matrices with entirely filled pores. The static capacitance of such samples evidently demonstrates anomalous behaviour with a precursor behaviour caused by orientational molecular ordering due to the paranematic-to-nematic transition. For the entire pore filling it can be well described by a phenomenological Landau-de Gennes theory (KKLZ approach).  The corresponding analysis yields the nematic ordering field, $\sigma$, equals 0.91 (5CB-$p$SiO$_2$) and 1.10 (6CB-$p$SiO$_2$) which in both cases corresponds to a supercritical regime and appears in good agreement with recent optical birefringence studies \cite{Calus2012}. Whereas the changes of static capacitance characterises the behaviour of the effective (averaged) order parameter over the pore filling, analysis of the relaxation strengths of the slow and fast relaxation processes provide an access to local molecular ordering in different parts of the pore filling. 

Thus, the molecular ordering in the surface or interface layer next to the pore walls and in the core region can be resolved. The molecular ordering in the core region is reminiscent of that in the bulk. It exhibits a strong increase in the nematic ordering just below the paranematic-to-nematic transition point, $T_{PN}$, and subsequent saturation at cooling. The surface ordering, on the other hand, evidently exhibits a continuous evolution with a smeared kink near $T_{PN}$ and asymptotic decreasing above this temperature. The molecular ordering in the core region of the pore filling influences the peripheral molecular layers, presumably via intermolecular interactions.

Overall, the dielectric spectroscopy study on the calamatic nematic crystals 5CB and 6CB confined to cylindrical nanochannels of monolithic silica membranes reveals features analogous to those found for 7CB in nanoporous silica substrates \cite{Calus2015}. Therefore, we believe that the dielectric spectroscopy technique along with the phenomenological approach and key physical principles outlined here and in Ref. \cite{Calus2015} are applicable to the important class of the cyanobiphenyl family and confined nematics in general.

An inhomogeneous orientational behaviour has also been found in Molecular Dynamics simulations for rod-like LCs interacting with Gay-Berne potentials for confinement in nanochannels \cite{Ji2009, Guegan2007, Lefort2008} and in slit-pore geometry \cite{Gruhn1997, Gruhn1998, Barmes2004}. The qualitative results of these studies agree with the findings presented here. Unfortunately, no partial fillings, in particular monolayer behavior, have been explored in the simulation studies. This may be an interesting task for simulation studies in the future. 

Moreover, in agreement with experimental studies \cite{Kityk2008, Kityk2010, Gear2015} molecular dynamics studies suggest a significant influence of interface roughness and/or microstructure on the anchoring strength and thus on the nematic-isotropic transition \cite{Cheung2006, Roscioni2013}, an interplay which may be explorable in the near future given the availability of hierarchically tailorable porous solids \cite{Kuester2014}.



\section{Acknowledgement}

This work has been supported by the Polish National Science Centre (NCN) under the Project "Molecular Structure and Dynamics of Liquid Crystals Based Nanocomposites" (Decision No. DEC-2012/05/B/ST3/02782). The German research foundation (DFG) funded the research by the research Grant No. Hu850/3 and within the collaborative research initiative "Tailor-made Multi-Scale Materials Systems" (SFB 986, project area B and project C2), Hamburg.

\footnotesize{
\bibliographystyle{rsc} 

\providecommand*{\mcitethebibliography}{\thebibliography}
\csname @ifundefined\endcsname{endmcitethebibliography}
{\let\endmcitethebibliography\endthebibliography}{}
}
\end{document}